\documentclass[twoside,reqno]{bjp}
\usepackage{epsfig,cite}
\usepackage{graphicx}
\usepackage{amssymb,amsmath}
\usepackage{times}
\begin{document}

\sloppy \raggedbottom
\setcounter{page}{1}


%
%
%
%
\newcommand\rf[1]{(\ref{eq:#1})}
\newcommand\lab[1]{\label{eq:#1}}
\newcommand\nonu{\nonumber}
\newcommand\br{\begin{eqnarray}}
\newcommand\er{\end{eqnarray}}
\newcommand\be{\begin{equation}}
\newcommand\ee{\end{equation}}
\newcommand\eq{\!\!\!\! &=& \!\!\!\! }
\newcommand\foot[1]{\footnotemark\footnotetext{#1}}
\newcommand\lb{\lbrack}
\newcommand\rb{\rbrack}
\newcommand\llangle{\left\langle}
\newcommand\rrangle{\right\rangle}
\newcommand\blangle{\Bigl\langle}
\newcommand\brangle{\Bigr\rangle}
\newcommand\llb{\left\lbrack}
\newcommand\rrb{\right\rbrack}
\newcommand\Blb{\Bigl\lbrack}
\newcommand\Brb{\Bigr\rbrack}
\newcommand\lcurl{\left\{}
\newcommand\rcurl{\right\}}
\renewcommand\({\left(}
\renewcommand\){\right)}
\renewcommand\v{\vert}                     
\newcommand\bv{\bigm\vert}               
\newcommand\Bgv{\;\Bigg\vert}            
\newcommand\bgv{\bigg\vert}              
\newcommand\lskip{\vskip\baselineskip\vskip-\parskip\noindent}
\newcommand\mskp{\par\vskip 0.3cm \par\noindent}
\newcommand\sskp{\par\vskip 0.15cm \par\noindent}
\newcommand\bc{\begin{center}}
\newcommand\ec{\end{center}}
\newcommand\Lbf[1]{{\Large \textbf{{#1}}}}
\newcommand\lbf[1]{{\large \textbf{{#1}}}}




\newcommand\tr{\mathop{\mathrm tr}}                  
\newcommand\Tr{\mathop{\mathrm Tr}}                  
\newcommand\partder[2]{\frac{{\partial {#1}}}{{\partial {#2}}}}
\newcommand\partderd[2]{{{\partial^2 {#1}}\over{{\partial {#2}}^2}}}
\newcommand\partderh[3]{{{\partial^{#3} {#1}}\over{{\partial {#2}}^{#3}}}}
\newcommand\partderm[3]{{{\partial^2 {#1}}\over{\partial {#2} \partial{#3} }}}
\newcommand\partderM[6]{{{\partial^{#2} {#1}}\over{{\partial {#3}}^{#4}{\partial {#5}}^{#6} }}}          
\newcommand\funcder[2]{{{\delta {#1}}\over{\delta {#2}}}}
\newcommand\Bil[2]{\Bigl\langle {#1} \Bigg\vert {#2} \Bigr\rangle}  
\newcommand\bil[2]{\left\langle {#1} \bigg\vert {#2} \right\rangle} 
\newcommand\me[2]{\left\langle {#1}\right|\left. {#2} \right\rangle} 

\newcommand\sbr[2]{\left\lbrack\,{#1}\, ,\,{#2}\,\right\rbrack} 
\newcommand\Sbr[2]{\Bigl\lbrack\,{#1}\, ,\,{#2}\,\Bigr\rbrack}
\newcommand\Gbr[2]{\Bigl\lbrack\,{#1}\, ,\,{#2}\,\Bigr\} }
\newcommand\pbr[2]{\{\,{#1}\, ,\,{#2}\,\}}       
\newcommand\Pbr[2]{\Bigl\{ \,{#1}\, ,\,{#2}\,\Bigr\}}  
\newcommand\pbbr[2]{\lcurl\,{#1}\, ,\,{#2}\,\rcurl}




\renewcommand\a{\alpha}
\renewcommand\b{\beta}
\renewcommand\c{\chi}
\renewcommand\d{\delta}
\newcommand\D{\Delta}
\newcommand\eps{\epsilon}
\newcommand\vareps{\varepsilon}
\newcommand\g{\gamma}
\newcommand\G{\Gamma}
\newcommand\grad{\nabla}
\newcommand\h{\frac{1}{2}}
\renewcommand\k{\kappa}
\renewcommand\l{\lambda}
\renewcommand\L{\Lambda}
\newcommand\m{\mu}
\newcommand\n{\nu}
\newcommand\om{\omega}
\renewcommand\O{\Omega}
\newcommand\p{\phi}
\newcommand\vp{\varphi}
\renewcommand\P{\Phi}
\newcommand\pa{\partial}
\newcommand\tpa{{\tilde \partial}}
\newcommand\bpa{{\bar \partial}}
\newcommand\pr{\prime}
\newcommand\ra{\rightarrow}
\newcommand\lra{\longrightarrow}
\renewcommand\r{\rho}
\newcommand\s{\sigma}
\renewcommand\S{\Sigma}
\renewcommand\t{\tau}
\renewcommand\th{\theta}
\newcommand\bth{{\bar \theta}}
\newcommand\Th{\Theta}
\newcommand\z{\zeta}
\newcommand\ti{\tilde}
\newcommand\wti{\widetilde}
\newcommand\twomat[4]{\left(\begin{array}{cc}  
{#1} & {#2} \\ {#3} & {#4} \end{array} \right)}
\newcommand\threemat[9]{\left(\begin{array}{ccc}  
{#1} & {#2} & {#3}\\ {#4} & {#5} & {#6}\\
{#7} & {#8} & {#9} \end{array} \right)}


\newcommand\cA{{\mathcal A}}
\newcommand\cB{{\mathcal B}}
\newcommand\cC{{\mathcal C}}
\newcommand\cD{{\mathcal D}}
\newcommand\cE{{\mathcal E}}
\newcommand\cF{{\mathcal F}}
\newcommand\cG{{\mathcal G}}
\newcommand\cH{{\mathcal H}}
\newcommand\cI{{\mathcal I}}
\newcommand\cJ{{\mathcal J}}
\newcommand\cK{{\mathcal K}}
\newcommand\cL{{\mathcal L}}
\newcommand\cM{{\mathcal M}}
\newcommand\cN{{\mathcal N}}
\newcommand\cO{{\mathcal O}}
\newcommand\cP{{\mathcal P}}
\newcommand\cQ{{\mathcal Q}}
\newcommand\cR{{\mathcal R}}
\newcommand\cS{{\mathcal S}}
\newcommand\cT{{\mathcal T}}
\newcommand\cU{{\mathcal U}}
\newcommand\cV{{\mathcal V}}
\newcommand\cX{{\mathcal X}}
\newcommand\cW{{\mathcal W}}
\newcommand\cY{{\mathcal Y}}
\newcommand\cZ{{\mathcal Z}}

\newcommand{\nit}{\noindent}
\newcommand{\ct}[1]{\cite{#1}}
\newcommand{\bib}[1]{\bibitem{#1}}

\newcommand\PRL[3]{\textsl{Phys. Rev. Lett.} \textbf{#1} (#2) #3}
\newcommand\NPB[3]{\textsl{Nucl. Phys.} \textbf{B#1} (#2) #3}
\newcommand\NPBFS[4]{\textsl{Nucl. Phys.} \textbf{B#2} [FS#1] (#3) #4}
\newcommand\CMP[3]{\textsl{Commun. Math. Phys.} \textbf{#1} (#2) #3}
\newcommand\PRD[3]{\textsl{Phys. Rev.} \textbf{D#1} (#2) #3}
\newcommand\PLA[3]{\textsl{Phys. Lett.} \textbf{#1A} (#2) #3}
\newcommand\PLB[3]{\textsl{Phys. Lett.} \textbf{#1B} (#2) #3}
\newcommand\CQG[3]{\textsl{Class. Quantum Grav.} \textbf{#1} (#2) #3}
\newcommand\JMP[3]{\textsl{J. Math. Phys.} \textbf{#1} (#2) #3}
\newcommand\PTP[3]{\textsl{Prog. Theor. Phys.} \textbf{#1} (#2) #3}
\newcommand\SPTP[3]{\textsl{Suppl. Prog. Theor. Phys.} \textbf{#1} (#2) #3}
\newcommand\AoP[3]{\textsl{Ann. of Phys.} \textbf{#1} (#2) #3}
\newcommand\RMP[3]{\textsl{Rev. Mod. Phys.} \textbf{#1} (#2) #3}
\newcommand\PR[3]{\textsl{Phys. Reports} \textbf{#1} (#2) #3}
\newcommand\FAP[3]{\textsl{Funkt. Anal. Prilozheniya} \textbf{#1} (#2) #3}
\newcommand\FAaIA[3]{\textsl{Funct. Anal. Appl.} \textbf{#1} (#2) #3}
\newcommand\TAMS[3]{\textsl{Trans. Am. Math. Soc.} \textbf{#1} (#2) #3}
\newcommand\InvM[3]{\textsl{Invent. Math.} \textbf{#1} (#2) #3}
\newcommand\AdM[3]{\textsl{Advances in Math.} \textbf{#1} (#2) #3}
\newcommand\PNAS[3]{\textsl{Proc. Natl. Acad. Sci. USA} \textbf{#1} (#2) #3}
\newcommand\LMP[3]{\textsl{Letters in Math. Phys.} \textbf{#1} (#2) #3}
\newcommand\IJMPA[3]{\textsl{Int. J. Mod. Phys.} \textbf{A#1} (#2) #3}
\newcommand\IJMPD[3]{\textsl{Int. J. Mod. Phys.} \textbf{D#1} (#2) #3}
\newcommand\TMP[3]{\textsl{Theor. Math. Phys.} \textbf{#1} (#2) #3}
\newcommand\JPA[3]{\textsl{J. Physics} \textbf{A#1} (#2) #3}
\newcommand\JSM[3]{\textsl{J. Soviet Math.} \textbf{#1} (#2) #3}
\newcommand\MPLA[3]{\textsl{Mod. Phys. Lett.} \textbf{A#1} (#2) #3}
\newcommand\JETP[3]{\textsl{Sov. Phys. JETP} \textbf{#1} (#2) #3}
\newcommand\JETPL[3]{\textsl{ Sov. Phys. JETP Lett.} \textbf{#1} (#2) #3}
\newcommand\PHSA[3]{\textsl{Physica} \textbf{A#1} (#2) #3}
\newcommand\PHSD[3]{\textsl{Physica} \textbf{D#1} (#2) #3}
\newcommand\JPSJ[3]{\textsl{J. Phys. Soc. Jpn.} \textbf{#1} (#2) #3}
\newcommand\JGP[3]{\textsl{J. Geom. Phys.} \textbf{#1} (#2) #3}

\newcommand\Xdot{\stackrel{.}{X}}
\newcommand\xdot{\stackrel{.}{x}}
\newcommand\ydot{\stackrel{.}{y}}
\newcommand\yddot{\stackrel{..}{y}}
\newcommand\rdot{\stackrel{.}{r}}
\newcommand\rddot{\stackrel{..}{r}}
\newcommand\vpdot{\stackrel{.}{\varphi}}
\newcommand\vpddot{\stackrel{..}{\varphi}}
\newcommand\tdot{\stackrel{.}{t}}
\newcommand\zdot{\stackrel{.}{z}}
\newcommand\etadot{\stackrel{.}{\eta}}
\newcommand\udot{\stackrel{.}{u}}
\newcommand\vdot{\stackrel{.}{v}}
\newcommand\rhodot{\stackrel{.}{\rho}}
\newcommand\xdotdot{\stackrel{..}{x}}
\newcommand\ydotdot{\stackrel{..}{y}}
\newcommand\adot{\stackrel{.}{a}}
\newcommand\addot{\stackrel{..}{a}}
\newcommand\Adot{\stackrel{.}{A}}
\newcommand\Bdot{\stackrel{.}{B}}
\newcommand\Hdot{\stackrel{.}{H}}



\title{Metric-Independent Volume-Forms in Gravity and Cosmology
\thanks{Invited talk at the Memorial {\em ``Matey Mateev Symposium''}, 
April 2015, \textsl{https://indico.cern.ch/event/359644/}. E.N. and S.P.
express their sincere gratitude and appreciation to the late Acad. Prof. M. Mateev -- 
one of their first university teachers.}}

\begin{start}
\author{E.~Guendelman}{1}, \coauthor{E.~Nissimov}{2},
\coauthor{S.~Pacheva}{2}

\address{Department of Physics, Ben-Gurion Univ. of the Negev,
Beer-Sheva 84105, Israel}{1}

\address{Institute of Nuclear Research and Nuclear Energy,
Bulg. Acad. Sci., Sofia 1784, Bulgaria}{2}

\runningheads{E.~Guendelman, E.~Nissimov, S.~Pacheva}{Metric-Independent Volume-Forms 
in Gravity and Cosmology}

\received{}


\begin{Abstract}
Employing alternative spacetime volume-forms (generally-covariant integration measure 
densities) independent of the pertinent Riemannian spacetime metric have profound 
impact in general relativity. Although formally appearing as ``pure-gauge'' 
dynamical degrees of freedom they trigger a number of remarkable physically 
important phenomena such as: (i) new mechanism of dynamical generation of 
cosmological constant; (ii) new type of ``quintessential inflation'' scenario in 
cosmology; (iii) non-singular initial ``emergent universe'' phase of
cosmological evolution preceding the inflationary phase;
(iv) new mechanism of dynamical spontaneous breakdown of supersymmetry 
in supergravity; (v) gravitational electrovacuum ``bags''. We study in
some detail the properties, together with their canonical Hamiltonian formulation, of a 
class of generalized gravity-matter models built with two independent
non-Riemannian volume-forms and discuss their implications in cosmology.
\end{Abstract}

\end{start}

\section[]{Introduction} 

Alternative spacetime volume-forms (generally-covariant integration measure densities) 
independent on the Riemannian metric on the pertinent spacetime manifold  have profound
impact in field theory models with general coordinate reparametrization
invariance -- general relativity and its extensions, strings and (higher-dimensional) 
membranes.

Although formally appearing as ``pure-gauge'' dynamical degrees of freedom 
the non-Riemannian volume-form fields trigger a number of remarkable physically important 
phenomena. Among the principal new phenomena are:
\begin{itemize}
\item
(i) new mechanism of dynamical generation of cosmological constant; 
\item
(ii) new type of "quintessential inflation" scenario in cosmology describing
both the ``early'' and ``late'' universe in terms of a single scalar field; 
\item
(iii) non-singular initial phase of cosmological evolution -- 
``no Big-Bang'' ``emergent universe'' -- preceding the inflationary phase;
\item
(iv) new mechanism of dynamical spontaneous breakdown of supersymmetry in supergravity; 
\item
(v) Coupling of non-Riemannian volume-form gravity-matter theories to a special 
non-standard kind of nonlinear gauge system containing the square-root of standard 
Maxwell/Yang-Mills Lagrangian yields charge confinement/deconfinement phases 
associated with gravitational electrovacuum ``bags''.
\end{itemize}

Properties (i)-(iii) are discussed in more details in what follows.

The ideas of the present formalism rely substantially on a series of previous papers
\ct{TMT-orig} (for recent developments, see Refs.\ct{TMT-recent}), where
a new class of generally-covariant (non-supersymmetric)
field theory models including gravity -- called ``two-measure theories''
(TMT) was proposed based on the principal proposal to employ an alternative volume 
form (volume element or generally-covariant integration measure) on the spacetime 
manifold in the pertinent Lagrangian action. 
TMT appear to be promising candidates for resolution of various problems in
modern cosmology: the {\em dark energy} and {\em dark matter} problems, the fifth 
force problem, etc. 

In standard generally-covariant theories (with action $S=\int d^D\! x \sqrt{-g} \cL$)
the Riemannian spacetime volume-form, \textsl{i.e.}, the integration measure density is
given by $\sqrt{-g}$, where $g \equiv \det\Vert g_{\m\n}\Vert$ is the determinant of the 
corresponding Riemannian metric $g_{\m\n}$.
$\sqrt{-g}$ transforms as scalar density under general coordinate reparametrizations.

There is {\em no a priori} any obstacle to employ instead of $\sqrt{-g}$ another
alternative non-Riemannian volume element given by the following {\em non-Riemannian} 
integration measure density:
\be
\Phi(B) \equiv \frac{1}{(D-1)!}\vareps^{\m_1\ldots\m_D}\, \pa_{\m_1} B_{\m_2\ldots\m_D}
\; .
\lab{Phi-D}
\ee
Here $B_{\m_1\ldots\m_{D-1}}$ is an auxiliary rank $(D-1)$ antisymmetric tensor gauge
field, which will turn out to be pure-gauge degree of freedom. 
$\Phi(B)$ -- the dual field-strength of $B_{\m_1\ldots\m_{D-1}}$ -- similarly transforms
as scalar density under general coordinate reparametrizations like $\sqrt{-g}$.

An important property of the present formalism is that the non-Riemannian measure density
$\Phi(B)$ becomes {\em on-shell} proportional to the standard Riemannian one $\sqrt{-g}$
(see Eq.\rf{bar-g} below), \textsl{i.e.}, the physical meaning of $\Phi(B)$ as a
measure is preserved.

In the next Section 2 we describe in some detail the construction within the
Lagrangian formalism of a new class of generalized gravity-matter theories
built in terms of two different non-Riemannian volume-forms and derive the
effective Lagrangian in the physical ``Einstein frame''. In Section 3 we
provide a general canonical Hamiltonian treatment of gravity-matter theories with
non-Riemannian volume-forms and elucidate the physical meaning of the
auxiliary volume-form fields. Section 4 is devoted to discussion of the
cosmological implications of the above class of generalized gravity-matter
theories with two non-Riemannian volume-forms. The central result here is
the derivation of an effective scalar (``inflaton'') potential with {\em two
infinitely large flat regions} with vastly different energy scales. In
Section 5 we briefly describe the construction of the ``emergent universe'' 
solution as a non-singular (``no Big-Bang'') initial phase of cosmological
evolution preceding the inflationary phase.

\section[]{Gravity-Matter Theories with Two Non-Riemannian Volume-Forms}

Let us now consider modified-measure gravity-matter theories constructed in
terms of two different non-Riemannian volume-forms (employing first-order Palatini 
formalism, and using units where $G_{\rm Newton} = 1/16\pi$)
\ct{emergent-quintess}:
\be
S = \int d^4 x\,\P_1 (A) \Bigl\lb R + L^{(1)} \Bigr\rb +  
\int d^4 x\,\P_2 (B) \Bigl\lb L^{(2)} + \eps R^2 + 
\frac{\P (H)}{\sqrt{-g}}\Bigr\rb \; .
\lab{TMMT}
\ee
Here and below the following notations are used:
\begin{itemize}
\item
$\P_{1}(A)$ and $\P_2 (B)$ are two independent non-Riemannian volume-forms:
\be
\P_1 (A) = \frac{1}{3!}\vareps^{\m\n\k\l} \pa_\m A_{\n\k\l} \quad ,\quad
\P_2 (B) = \frac{1}{3!}\vareps^{\m\n\k\l} \pa_\m B_{\n\k\l} \; .
\lab{Phi-1-2}
\ee
\item
$\P (H)$ is the dual field-strength of a third auxiliary gauge field $H_{\m\n\l}$:
\be
\P (H) = \frac{1}{3!}\vareps^{\m\n\k\l} \pa_\m H_{\n\k\l} \; ,
\lab{Phi-H}
\ee
whose presence is essential for the consistency of \rf{TMMT}.
\item
$R = g^{\m\n} R_{\m\n}(\G)$ and $R_{\m\n}(\G)$ are the scalar curvature and the 
Ricci tensor in the first-order (Palatini) formalism, where the affine
connection $\G^\m_{\n\l}$ is \textsl{a priori} independent of the metric $g_{\m\n}$.
In the second action term in \rf{TMMT} we have added a $R^2$ gravity term
(again in the Palatini form). The gravity model $R+R^2$ within the
second order formalism was the first inflationary model originally
proposed in Ref.\ct{starobinsky}.
\item
$L^{(1,2)}$ denote two different Lagrangians of a single scalar matter field 
(``dilaton'' or ``inflaton'') of the form:
\br
L^{(1)} = -\h g^{\m\n} \pa_\m \vp \pa_\n \vp - V(\vp) \quad ,\quad
V(\vp) = f_1 \exp \{-\a\vp\} \; ,
\lab{L-1} \\
L^{(2)} = -\frac{b}{2} e^{-\a\vp} g^{\m\n} \pa_\m \vp \pa_\n \vp + U(\vp) 
\quad ,\quad U(\vp) = f_2 \exp \{-2\a\vp\} \; ,
\lab{L-2}
\er
where $\a, f_1, f_2$ are dimensionful positive parameters, whereas $b$ is a
dimensionless one.
\end{itemize}

The action \rf{TMMT} possesses a {\em global Weyl-scale invariance}:
\br
g_{\m\n} \to \l g_{\m\n} \;,\; \G^\m_{\n\l} \to \G^\m_{\n\l} \; ,\; 
\vp \to \vp + \frac{1}{\a}\ln\l \; ,
\lab{scale-transf} \\
A_{\m\n\k} \to \l A_{\m\n\k} \; ,\; B_{\m\n\k} \to \l^2 B_{\m\n\k} \; ,\; 
H_{\m\n\k} \to H_{\m\n\k} \; .
\nonu
\er

The equations of motion w.r.t. affine connection $\G^\m_{\n\l}$ yield the
following solution for the latter as a Levi-Civita connection:
\be
\G^\m_{\n\l} = \G^\m_{\n\l}({\bar g}) = 
\h {\bar g}^{\m\k}\(\pa_\n {\bar g}_{\l\k} + \pa_\l {\bar g}_{\n\k} 
- \pa_\k {\bar g}_{\n\l}\) \; ,
\lab{G-eq}
\ee
corresponding to the Weyl-rescaled metric ${\bar g}_{\m\n}$:
\be
{\bar g}_{\m\n} = (\chi_1 + 2\eps \chi_2 R) g_{\m\n} \;\; ,\;\; 
\chi_1 \equiv \frac{\P_1 (A)}{\sqrt{-g}} \;\; ,\;\;
\chi_2 \equiv \frac{\P_2 (B)}{\sqrt{-g}} \; .
\lab{bar-g}
\ee
Transition from the original metric $g_{\m\n}$ to ${\bar g}_{\m\n}$ realizes
the passage to the physical {\em ``Einstein-frame''}, where the gravity equations
of motion acquire the standard form of Einstein's equations:
\be
R_{\m\n}({\bar g}) - \h {\bar g}_{\m\n} R({\bar g}) = \h T^{\rm eff}_{\m\n}
\lab{einstein-frame-eqs}
\ee
with an appropriate {\em effective matter energy-momentum tensor} given in terms
of an {\em effective Einstein-frame matter Lagrangian} $L_{\rm eff}$ 
(see \rf{L-eff-final} below).

Variation of the action \rf{TMMT} w.r.t. auxiliary tensor gauge fields
$A_{\m\n\l}$, $B_{\m\n\l}$ and $H_{\m\n\l}$ yields the equations:
\be
\pa_\m \Bigl\lb R + L^{(1)} \Bigr\rb = 0 \;, \;
\pa_\m \Bigl\lb L^{(2)} + \eps R^2 + \frac{\P (H)}{\sqrt{-g}}\Bigr\rb = 0 
\;, \; \pa_\m \Bigl(\frac{\P_2 (B)}{\sqrt{-g}}\Bigr) = 0 \; ,
\lab{A-B-H-eqs}
\ee
whose solutions read:
\br
\frac{\P_2 (B)}{\sqrt{-g}} \equiv \chi_2 = {\rm const} \;\; ,\;\;
R + L^{(1)} = - M_1 = {\rm const} \; ,
\nonu \\
L^{(2)} + \eps R^2 + \frac{\P (H)}{\sqrt{-g}} = - M_2  = {\rm const} \; .
\lab{integr-const}
\er
Here $M_1$ and $M_2$ are arbitrary dimensionful and $\chi_2$
arbitrary dimensionless integration constants.

The first integration constant $\chi_2$ in \rf{integr-const} preserves
global Weyl-scale invariance \rf{scale-transf}
whereas the appearance of the second and third integration constants $M_1,\, M_2$
signifies {\em dynamical spontaneous breakdown} of global Weyl-scale invariance 
under \rf{scale-transf} 
due to the scale non-invariant solutions (second and third ones) in \rf{integr-const}. 

It is very instructive to elucidate the physical meaning of the three arbitrary 
integration constants $M_1,\, M_2,\,\chi_2$ from the point of view of the
canonical Hamiltonian formalism. Namely,  
$M_1,\, M_2,\,\chi_2$ are identified as conserved Dirac-constrained
canonical momenta conjugated to (certain components of) the auxiliary
maximal rank antisymmetric tensor gauge fields $A_{\m\n\l}, B_{\m\n\l}, H _{\m\n\l}$
entering the original non-Riemannian volume-form action \rf{TMMT} 
(for details, see next Section 3 below).

Varying \rf{TMMT} w.r.t. $g_{\m\n}$ and using relations \rf{integr-const} 
we have:
\be
\chi_1 \Bigl\lb R_{\m\n} + \h\( g_{\m\n}L^{(1)} - T^{(1)}_{\m\n}\)\Bigr\rb -
\h \chi_2 \Bigl\lb T^{(2)}_{\m\n} + g_{\m\n} \(\eps R^2 + M_2\)
- 2 R\,R_{\m\n}\Bigr\rb = 0 \; ,
\lab{pre-einstein-eqs}
\ee
where $\chi_1$ and $\chi_2$ are defined in \rf{bar-g},
and $T^{(1,2)}_{\m\n}$ are the energy-momentum tensors of the scalar
field Lagrangians with the standard definitions:
\be
T^{(1,2)}_{\m\n} = g_{\m\n} L^{(1,2)} - 2 \partder{}{g^{\m\n}} L^{(1,2)} \; .
\lab{EM-tensor}
\ee

Taking the trace of Eqs.\rf{pre-einstein-eqs} and using again second relation 
\rf{integr-const} we solve for the scale factor $\chi_1$:
\be
\chi_1 = 2 \chi_2 \frac{T^{(2)}/4 + M_2}{L^{(1)} - T^{(1)}/2 - M_1} \; ,
\lab{chi-1}
\ee
where $T^{(1,2)} = g^{\m\n} T^{(1,2)}_{\m\n}$. 

Using second relation \rf{integr-const} Eqs.\rf{pre-einstein-eqs} can be put 
in the Einstein-like form:
\br
R_{\m\n} - \h g_{\m\n}R = \h g_{\m\n}\(L^{(1)} + M_1\)
+ \frac{1}{2\O}\(T^{(1)}_{\m\n} - g_{\m\n}L^{(1)}\)
\nonu \\
+ \frac{\chi_2}{2\chi_1 \O} \Bigl\lb T^{(2)}_{\m\n} + 
g_{\m\n} \(M_2 + \eps(L^{(1)} + M_1)^2\)\Bigr\rb \; ,
\lab{einstein-like-eqs}
\er
where:
\be
\O = 1 - \frac{\chi_2}{\chi_1}\,2\eps\(L^{(1)} + M_1\) \; .
\lab{Omega-eq}
\ee
Let us note that \rf{bar-g}, upon taking into account second relation
\rf{integr-const} and \rf{Omega-eq}, can be written as:
\be
{\bar g}_{\m\n} = \chi_1\O\,g_{\m\n} \; .
\lab{bar-g-2}
\ee

Now, we can bring Eqs.\rf{einstein-like-eqs} into the standard form of Einstein 
equations for the rescaled  metric ${\bar g}_{\m\n}$ \rf{bar-g-2}, 
\textsl{i.e.}, the Einstein-frame equations: 
\be
R_{\m\n}({\bar g}) - \h {\bar g}_{\m\n} R({\bar g}) = \h T^{\rm eff}_{\m\n}
\lab{eff-einstein-eqs}
\ee
with energy-momentum tensor corresponding according to the definition \rf{EM-tensor}:
\be
T^{\rm eff}_{\m\n} = g_{\m\n} L_{\rm eff} - 2 \partder{}{g^{\m\n}} L_{\rm eff}
\lab{T-eff}
\ee
to the following effective (Einstein-frame) scalar field Lagrangian 
of non-canonical ``k-essence'' (kinetic quintessence) type \ct{k-essence}
($X \equiv - \h {\bar g}^{\m\n} \pa_\m \vp \pa_\n \vp$ denotes the scalar kinetic term):
\be
L_{\rm eff} = A(\vp) X + B(\vp) X^2 - U_{\rm eff}(\vp) \; ,
\lab{L-eff-final}
\ee
where (recall $V=f_1 e^{-\a\vp}$ and $U=f_2 e^{-2\a\vp}$):
\br
A(\vp) \equiv 1 + \Bigl\lb \h b e^{-\a\vp} - \eps (V - M_1)\Bigr\rb
\frac{V - M_1}{U + M_2 + \eps (V - M_1)^2} \; ,
\lab{A-def} \\
B(\vp) \equiv \chi_2 \frac{\eps\Bigl\lb U + M_2 + (V - M_1) b e^{-\a\vp}\Bigr\rb
- \frac{1}{4} b^2 e^{-2\a\vp}}{U + M_2 + \eps (V - M_1)^2} \; ,
\lab{B-def}\\
U_{\rm eff} (\vp) \equiv 
\frac{(V - M_1)^2}{4\chi_2 \Bigl\lb U + M_2 + \eps (V - M_1)^2\Bigr\rb} \; .
\lab{U-eff}
\er

\section[]{Canonical Hamiltonian Treatment of Gravity-Matter Theories with
Non-Riemannian Volume-Forms}

Here we will briefly discuss the application of the canonical Hamiltonian formalism 
to the new gravity-matter model based on two non-Riemannian spacetime volume-forms 
\rf{TMMT}. In order to elucidate the proper physical meaning of the arbitrary
integration constants $\chi_2,\, M_1,\, M_2$ \rf{integr-const} encountered
within the Lagrangian formalism's treatment of \rf{TMMT} it is sufficient to
concentrate only on the canonical Hamiltonian structure related to the
auxiliary maximal rank antisymmetric tensor gauge fields 
$A_{\m\n\l}, B_{\m\n\l}, H_{\m\n\l}$ and their respective conjugate momenta.

For convenience let us introduce the following short-hand notations for the
field-strengths \rf{Phi-1-2}, \rf{Phi-H} of the auxiliary 3-index antisymmetric gauge 
fields (the dot indicating time-derivative): 
\br
\P_1 (A) = \Adot + \pa_i A^i \quad, \quad 
A = \frac{1}{3!} \vareps^{ijk} A_{ijk} \;\; ,\;\;
A^i = - \h \vareps^{ijk} A_{0jk} \; ,
\lab{A-can} \\
\P_2 (B) = \Bdot + \pa_i B^i \quad, \quad 
B = \frac{1}{3!} \vareps^{ijk} B_{ijk} \;\; ,\;\;
B^i = - \h \vareps^{ijk} B_{0jk} \; ,
\lab{B-can} \\
\P (H) = \Hdot + \pa_i H^i \quad, \quad 
H = \frac{1}{3!} \vareps^{ijk} H_{ijk} \;\; ,\;\;
H^i = - \h \vareps^{ijk} H_{0jk} \; ,
\lab{H-can}
\er
Also we will use the short-hand notation:
\be
{\wti L}^{(1)} (u,\udot) \equiv R + L^{(1)} \quad ,\quad
{\wti L}^{(2)} (u,\udot) \equiv L^{(2)} + \eps R^2 \; ,
\lab{L-tilde}
\ee
where $L^{(1,2)}$ are as in \rf{L-1}-\rf{L-2} and where $(u,\udot)$ collectively denote 
the set of the basic gravity-matter canonical variables 
$(u)=\bigl(g_{\m\n}, \vp, A_\m \bigr)$ and their respective velocities.

For the pertinent canonical momenta conjugated to \rf{A-can}-\rf{H-can} we have:
\br
\pi_A = {\wti L}_1 (u,\udot) \;\; ,\;\;
\pi_B = {\wti L}^{(2)} (u,\udot) + \frac{1}{\sqrt{-g}}(\Hdot + \pa_i H^i) \; ,
\nonu \\
\pi_H = \frac{1}{\sqrt{-g}}(\Bdot + \pa_i B^i) \; ,
\lab{can-momenta-aux}
\er
and:
\be
\pi_{A^i} = 0 \quad,\quad \pi_{B^i} = 0 \quad,\quad \pi_{H^i} = 0 \; .
\lab{can-momenta-zero}
\ee
The latter imply that $A^i, B^i, H^i$ will in fact appear as Lagrange multipliers
for certain first-class Hamiltonian constraints 
(see Eqs.\rf{pi-A-const}-\rf{pi-B-pi-H-const} below). 
For the canonical momenta conjugated to the basic gravity-matter canonical variables 
we have (using last relation \rf{can-momenta-aux}):
\be
p_u = (\Adot + \pa_i A^i) \frac{\pa}{\pa \udot} {\wti L}_1 (u,\udot) + 
\pi_H \sqrt{-g} \frac{\pa}{\pa \udot} L^{(2)} (u,\udot) \; .
\lab{can-momenta-u}
\ee

Now, relations \rf{can-momenta-aux} and \rf{can-momenta-u} allow us to
obtain the velocities $\udot,\,\Adot,\,\Bdot,\,\Hdot$ as functions
of the canonically conjugate momenta $\udot = \udot (u,p_u,\pi_A,\pi_B,\pi_H)$
\textsl{etc.} (modulo some Dirac constraints among the basic gravity-matter
variables due to general coordinate and gauge invariances). Taking into account
\rf{can-momenta-aux}-\rf{can-momenta-zero} (and the short-hand notations
\rf{A-can}-\rf{L-tilde}) the canonical Hamiltonian corresponding to \rf{TMMT}:
\br
\cH = p_u \udot + \pi_A \Adot + \pi_B \Bdot + \pi_H \Hdot -
(\Adot + \pa_i A^i) {\wti L}_1 (u,\udot) 
\nonu \\
- \pi_H \sqrt{-g} \Bigl\lb {\wti L}^{(2)}(u,\udot) + 
\frac{1}{\sqrt{-g}}(\Hdot + \pa_i H^i) \Bigr\rb
\lab{can-hamiltonian}
\er
acquires the following form as function of the canonically conjugated variables
(here $\udot = \udot (u,p_u,\pi_A,\pi_B,\pi_H)$):
\br
\cH = p_u \udot - \pi_H \sqrt{-g} {\wti L}^{(2)}(u,\udot)
\nonu \\
+ \sqrt{-g} \pi_H \pi_B - \pa_i A^i \pi_A - \pa_i B^i \pi_B - \pa_i H^i \pi_H \; .
\lab{can-hamiltonian-final}
\er
From \rf{can-hamiltonian-final} we deduce that indeed $A^i, B^i, H^i$ are Lagrange 
multipliers for the first-class Hamiltonian constraints:
\be
\pa_i \pi_A = 0 \;\; \to\;\; \pi_A = - M_1 = {\rm const} \; ,
\lab{pi-A-const}
\ee
and similarly:
\be
\pi_B = - M_2 = {\rm const} \quad ,\quad \pi_H = \chi_2 = {\rm const} \; ,
\lab{pi-B-pi-H-const}
\ee
which are the canonical Hamiltonian counterparts of Lagrangian constraint
equations of motion \rf{integr-const}.

Thus, the canonical Hamiltonian treatment of \rf{TMMT} reveals the meaning
of the auxiliary 3-index antisymmetric tensor gauge fields
$A_{\m\n\l},\, B_{\m\n\l},\, H_{\m\n\l}$ -- building blocks of
the non-Riemannian spacetime volume-form formulation of the modified gravity-matter
model \rf{TMMT}. Namely, the canonical momenta $\pi_A,\, \pi_B,\, \pi_H$ 
conjugated to the ``magnetic'' parts $A,B,H$ \rf{A-can}-\rf{H-can}
of the auxiliary 3-index antisymmetric tensor gauge fields are constrained
through Dirac first-class constraints \rf{pi-A-const}-\rf{pi-B-pi-H-const}
to be constants identified with the arbitrary 
integration constants $\chi_2,\, M_1,\, M_2$ \rf{integr-const} arising within the 
Lagrangian formulation of the model. The canonical momenta 
$\pi_A^i,\, \pi_B^i,\, \pi_H^i$ conjugated to the ``electric'' parts $A^i,B^i,H^i$ 
\rf{A-can}-\rf{H-can} of the auxiliary 3-index antisymmetric tensor gauge field
are vanishing \rf{can-momenta-zero} which makes the latter canonical Lagrange 
multipliers for the above Dirac first-class constraints.

\section[]{Implications for Cosmology}

The most remarkable feature of the effective scalar potential $U_{\rm eff}(\vp)$ 
\rf{U-eff} is the existence of the following two 
{\em infinitely large flat regions} as function of $\vp$:

\begin{itemize}
\item
{\em (-) flat region} -- for large negative values of $\vp$:
\be
U_{\rm eff}(\vp) \simeq U_{(-)} \equiv 
\frac{f_1^2/f_2}{4\chi_2 (1+\eps f_1^2/f_2)} \; ,
\lab{U-minus} 
\ee
\item
{\rm (+) flat region} -- for large positive values of $\vp$:
\be
U_{\rm eff}(\vp) \simeq U_{(+)} \equiv 
\frac{M_1^2/M_2}{4\chi_2 (1+\eps M_1^2/M_2)} \; ,
\lab{U-plus}
\ee
\end{itemize}

The qualitative shape of $U_{\rm eff}(\vp)$ \rf{U-eff} is depicted on Figs.1
and 2.

\begin{figure}
\begin{center}
\includegraphics[width=9cm,keepaspectratio=true]{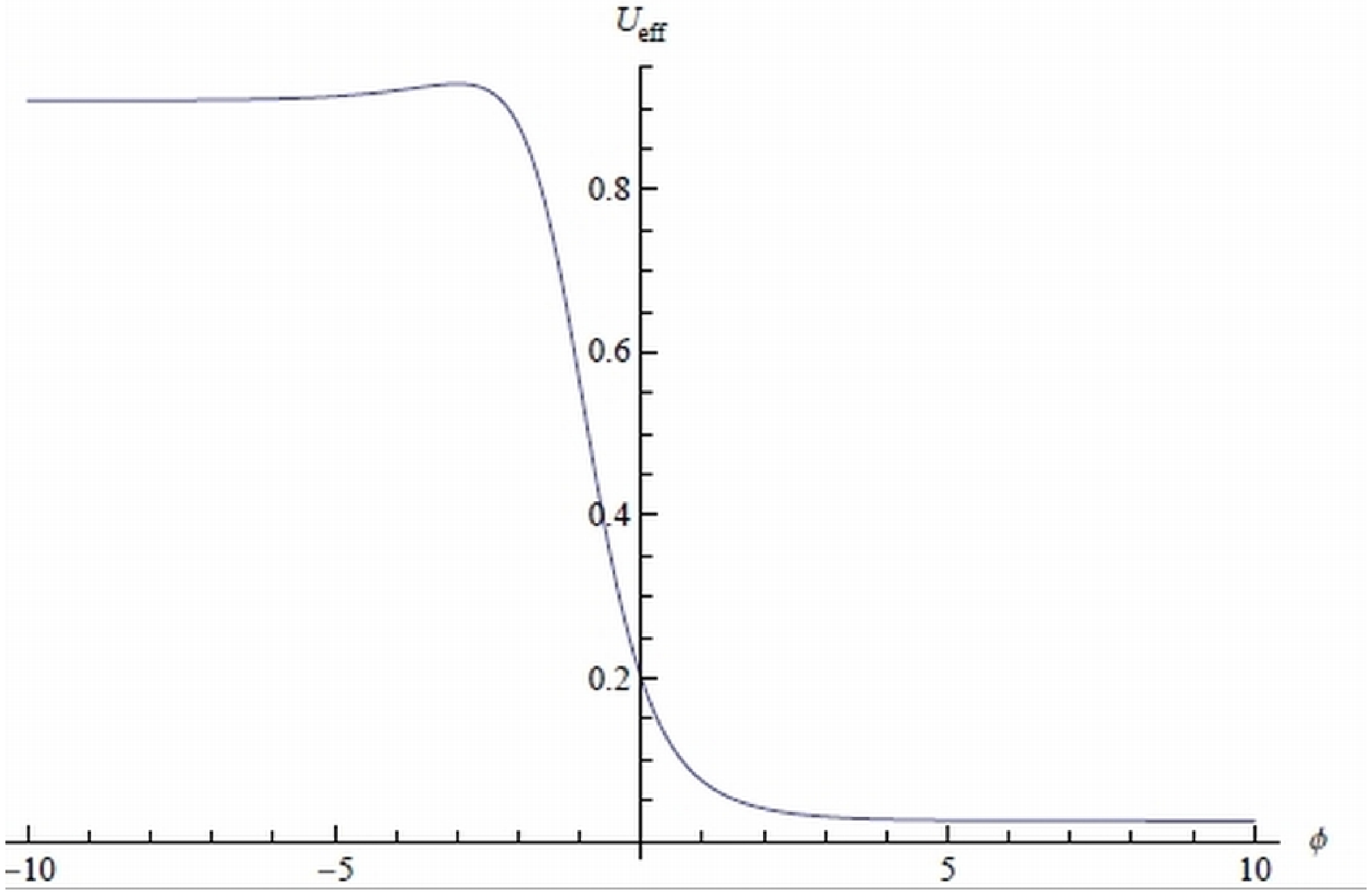}
\caption{Qualitative shape of the effective scalar potential $U_{\rm eff}$ \rf{U-eff}
as function of $\vp$ for $M_1 < 0$.}
\end{center}
\end{figure}

\begin{figure}
\begin{center}
\includegraphics[width=9cm,keepaspectratio=true]{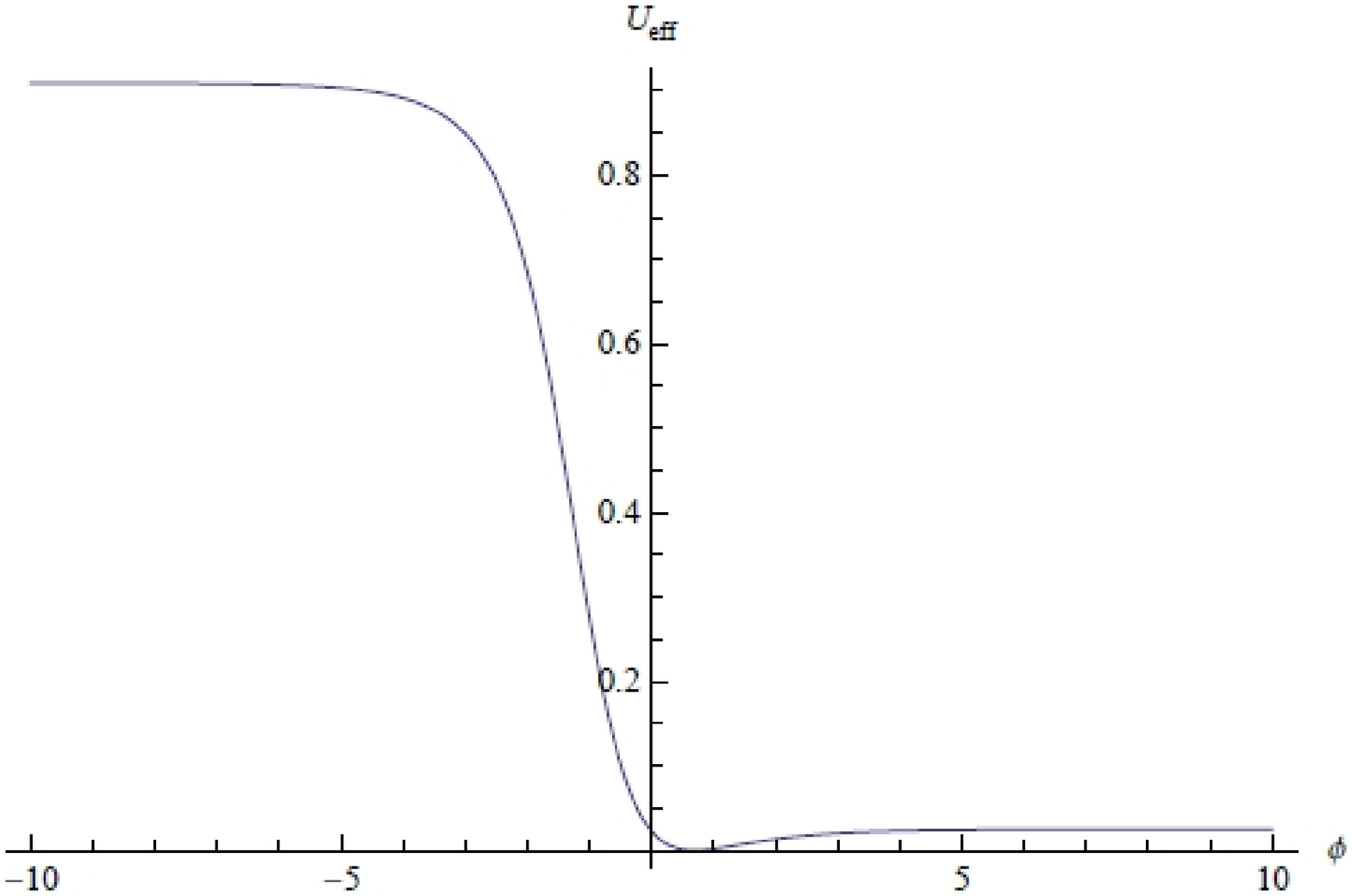}
\caption{Qualitative shape of the effective scalar potential $U_{\rm eff}$ \rf{U-eff}
as function of $\vp$ for $M_1 > 0$.}
\end{center}
\end{figure}

From the expression for $U_{\rm eff} (\vp)$ \rf{U-eff} and the Figures 1 and 2 we
deduce that we have an {\em explicit realization of quintessential inflation scenario}
\ct{quintess-basic}:
continuously connecting an inflationary phase of the universe's evolution
corresponding to the (most of the) $(-)$-flat region to a slowly accelerating
``present-day'' universe corresponding to the $(+)$-flat region through the 
evolution of a single scalar field.

The flat regions \rf{U-minus} and \rf{U-plus} correspond 
to the evolution of {\em early} and the {\em late} universe, respectively, 
provided we choose the ratio of the coupling constants in the original scalar potentials 
versus the ratio of the scale-symmetry breaking integration constants to obey
the following strong inequality:
\be
\frac{f_1^2/f_2}{1+\eps f_1^2/f_2} \gg \frac{M_1^2/M_2}{1+\eps M_1^2/M_2} \, ,
\lab{early-vs-late}
\ee
which makes the {\em vacuum energy density of the early universe} $U_{(-)}$ \rf{U-minus}
much bigger than that of the late universe $U_{(+)}$ \rf{U-plus}).

The inequality \rf{early-vs-late} is equivalent to the requirements:
\be
\frac{f_1^2}{f_2} \gg \frac{M_1^2}{M_2} \quad ,\quad |\eps| \frac{M_1^2}{M_2} \ll 1 \; .
\lab{early-vs-late-2}
\ee
If we choose the scales $|M_1| \sim M^4_{EW}$ and $M_2 \sim M^4_{Pl}$ 
\ct{arkani-hamed}, where $M_{EW},\, M_{Pl}$ are the electroweak and Planck scales, 
respectively, we are then naturally led to a very small vacuum energy density:
\be
U_{(+)}\sim M^8_{EW}/M^4_{Pl} \sim 10^{-120} M^4_{Pl} \; ,
\lab{U-plus-magnitude}
\ee
which is the right order of magnitude for the present epoch's vacuum energy density.

On the other hand, if we take the order of magnitude of the coupling constants 
in the effective potential $f_1 \sim f_2 \sim (10^{-2} M_{Pl})^4$, then the order of
magnitude of the vacuum energy density of the early universe becomes:
\be
U_{(-)} \sim f_1^2/f_2 \sim 10^{-8} M_{Pl}^4 \; ,
\lab{U-minus-magnitude}
\ee
which conforms to the Planck Collaboration data 
\ct{Planck} implying the energy scale of inflation to be of order $10^{-2} M_{Pl}$.

\section[]{``Emergent universe''}

Within the present gravity-matter theory with two non-Riemannian spacetime 
volume-forms we find explicit cosmological solution of the Einstein-frame system 
with effective scalar field Lagrangian \rf{L-eff-final}-\rf{U-eff} describing 
an epoch of a {\em non-singular creation of the universe} -- ``emergent universe''
\ct{emergent-univ}, preceding the inflationary phase.

The starting point are the Friedman equations \ct{weinberg-72}:
\be
\frac{\addot}{a}= - \frac{1}{12} (\rho + 3p) \quad ,\quad
H^2 + \frac{K}{a^2} = \frac{1}{6}\rho \quad ,\;\; H\equiv \frac{\adot}{a} \; ,
\lab{friedman-eqs}
\ee
describing the universe' evolution. Here:
\br
\rho = \h A(\vp) \vpdot^2 + \frac{3}{4} B(\vp) \vpdot^4 + U_{\rm eff}(\vp) \; ,
\lab{rho-def} \\
p = \h A(\vp) \vpdot^2 + \frac{1}{4} B(\vp) \vpdot^4 - U_{\rm eff}(\vp)
\lab{p-def}
\er
are the energy density and pressure of the scalar field $\vp = \vp (t)$, $H$
is the Hubble parameter and $K$ denotes the Gaussian curvature of the spacial section
in the Friedman-Lemaitre-Robertson-Walker metric \ct{weinberg-72}:
\be
ds^2 = - dt^2 + a^2(t) \Bigl\lb \frac{dr^2}{1-K r^2}
+ r^2 (d\th^2 + \sin^2\th d\phi^2)\Bigr\rb \; .
\lab{FLRW}
\ee

``Emergent universe'' is defined as a solution of the Friedman Eqs.\rf{friedman-eqs}
subject to the condition on the Hubble parameter $H$:
\be
H=0 \;\; \to \;\; a(t) = a_0 = {\rm const} \,,\;\; \rho + 3p =0 \;\; ,\;\;
\frac{K}{a_0^2} = \frac{1}{6}\rho ~(= {\rm const}) \; ,
\lab{emergent-cond}
\ee
with $\rho$ and $p$ as in \rf{rho-def}-\rf{p-def}. Here $K=1$ (``Einstein universe'').

The ``emergent universe'' condition \rf{emergent-cond} implies that the $\vp$-velocity
$\vpdot \equiv \vpdot_0$ is time-independent and satisfies the bi-quadratic 
algebraic equation:
\be
\frac{3}{2} B_{(-)}\vpdot_0^4 + 2 A_{(-)}\vpdot_0^2 - 2 U_{(-)} = 0 \; ,
\lab{vpdot-eq}
\ee
where $A_{(-)},\, B_{(-)},\, U_{(-)}$ are the limiting values on the $(-)$ flat region
of $A(\vp),\, B(\vp),\, U_{\rm eff}(\vp)$ \rf{A-def}-\rf{U-eff}.

The solution of Eq.\rf{vpdot-eq} reads:
\be
\vpdot_0^2 = - \frac{2}{3B_{(-)}} \Bigl\lb A_{(-)} \mp
\sqrt{A_{(-)}^2 + 3 B_{(-)}U_{(-)}}\Bigr\rb \; .
\lab{vpdot-sol}
\ee
and, thus, the ``emergent universe'' is characterized with {\em finite initial} 
Friedman factor and density:
\be
a_0^2 = \frac{6K}{\rho_0} \quad ,\quad
\rho_0 = \h A_{(-)}\vpdot_0^2 + \frac{3}{4} B_{(-)}\vpdot_0^4 + U_{(-)} \; ,
\lab{emergent-univ}
\ee
with $\vpdot_0^2$ as in \rf{vpdot-sol}.

Analysis of stability of the ``emergent universe'' solution \rf{emergent-univ} 
yields a harmonic oscillator type equation for the perturbation of the
Friedman factor $\d a$:
\be
\d \addot + \om^2 \d a = 0 \quad ,\quad
\om^2 \equiv \frac{2}{3}\rho_0\,\frac{\sqrt{A_{(-)}^2 + 3B_{(-)}U_{(-)}}}{A_{(-)} -
2\sqrt{A_{(-)}^2 + 3 B_{(-)}U_{(-)}}} \; .
\lab{stability-eq}
\ee
Thus stability condition $\om^2 >0$ leads to the following constraint on the coupling 
parameters:
\be
{\rm max} \Bigl\{-2\,,\, -8\bigl(1+3\eps f_1^2/f_2\bigr)
\Bigl\lb 1 - \sqrt{1 - \frac{1}{4\bigl(1+3\eps f_1^2/f_2\bigr)}}\Bigr\rb\Bigr\}
< b\frac{f_1}{f_2} < -1  \; .
\lab{param-constr}
\ee

Since the ratio $\frac{f_1^2}{f_2}$ proportional to the
height of the $(-)$ flat region of the effective scalar potential,
\textsl{i.e.}, the vacuum energy density in the early universe, must be
large (cf. \rf{early-vs-late}), we find that the lower end of the interval in 
\rf{param-constr} is very close to the upper end, \textsl{i.e.}, 
$b\frac{f_1}{f_2} \simeq -1$.

From Eqs.\rf{vpdot-sol}-\rf{emergent-univ} we obtain an inequality satisfied by the
initial energy density $\rho_0$ in the emergent universe:
\be
U_{(-)} < \rho_0 < 2U_{(-)} \; ,
\lab{rho-0}
\ee
which together with the estimate of the order of magnitude for 
$U_{(-)}$ \rf{U-minus-magnitude} implies order of magnitude for the initial
Friedman factor:
\be
a_0^2 \sim 10^{-8} K M_{Pl}^{-2} 
\lab{a-0}
\ee
(recall $K$ is the Gaussian curvature of the spacial section).

\section[]{Conclusions}
\begin{itemize}
\item
Non-Riemannian volume-form formalism in gravity/matter theories 
(\textsl{i.e.}, employing alternative non-Riemannian reparametrization covariant 
integration measure densities on the spacetime manifold) naturally generates a 
{\em dynamical cosmological constant} as an arbitrary dimensionful 
integration constant.
\item
Employing two different non-Riemannian volume-forms leads to the construction of a
new class of gravity-matter models, which produce an effective scalar potential with 
{\em two infinitely large flat regions}. This allows for a unified description of both 
early universe inflation as well as of present dark energy epoch.
\item
A remarkable feature is the existence of a stable initial phase of
{\em non-singular} universe creation preceding the inflationary phase
-- ``emergent universe'' without ``Big-Bang''.
\end{itemize}

Further very interesting features of gravity-matter theories built with
non-Riemannian spacetime volume-forms include:
\begin{itemize}
\item
Within non-Riemannian-modified-measure minimal $N=1$ supergravity the 
dynamically generated cosmological constant triggers spontaneous supersymmetry
breaking and mass generation for the gravitino (supersymmetric 
Brout-Englert-Higgs effect) \ct{susyssb}. Applying the same non-Riemannian
volume-form formalism to anti-de Sitter supergravity allows to produce
simultaneously a very large physical gravitino mass and a very small {\em
positive} observable cosmological constant  \ct{susyssb} in accordance with modern 
cosmological scenarios for slowly expanding universe of the present epoch
\ct{slow-accel}.
\item
Adding interaction with a special nonlinear (``square-root'' Maxwell) gauge field 
(known to describe charge confinement in flat spacetime) produces various
phases with different strength of confinement and/or with deconfinement, 
as well as gravitational electrovacuum ``bags'' partially mimicking the properties 
of {\em MIT bags} and solitonic constituent quark models (for details, see \ct{buggy}).
\end{itemize}

\section*{Acknowledgments}
We gratefully acknowledge support of our collaboration through the academic exchange 
agreement between the Ben-Gurion University and the Bulgarian Academy of Sciences.
E.N. and S.P. are supported by Bulgarian NSF Grant DFNI T02/6.
S.P. and E.N. have received partial support from COST actions MP-1210 and
MP-1405, respectively.


\end{document}